\documentclass[preprint,12pt]{elsarticle}

\usepackage{amssymb,amsmath}
\usepackage[colorlinks,bookmarksopen,bookmarksnumbered,citecolor=red,urlcolor=red]{hyperref}
\usepackage{pxfonts}
\usepackage[toc,page]{appendix}
\usepackage{lipsum}
\usepackage{hyperref}
\usepackage{cleveref}
\usepackage{multirow}

\numberwithin{equation}{section}
\newtheorem{theorem}{Theorem}[section]

\newtheorem{prop}[theorem]{Proposition}

\journal{impacted journal for publication}

\begin{document}

\begin{frontmatter}



\title{{\bf{On Poisson-exponential-Tweedie models for ultra-overdispersed data}}}
\author[label1]{Rahma \textsc{Abid}}\ead{Rahma.abid.ch@gmail.com}\address[label1]{Laboratory of Probability and Statistics.
Sfax Faculty of Sciences. B.P. 1171, CP 3038 Tunisia.}

\author[label2]{C\'{e}lestin C. \textsc{Kokonendji}}\ead{celestin.kokonendji@univ-fcomte.fr}\address[label2]{Laboratoire de Math\'{e}matiques de Besan\c{c}on, Universit\'{e} Bourgogne Franche-Comt\'{e}, France.}
\author[label1]{Afif \textsc{Masmoudi}}\ead{Afif.Masmoudi@fss.rnu.tn}

\address{}

\begin{abstract}
We introduce a new class of Poisson-exponential-Tweedie (PET) mixture in the framework of generalized linear models for ultra-overdispersed count data. The mean-variance relationship is of the form $m+m^{2}+\phi m^{p}$, where $\phi$ and $p$ are the dispersion and Tweedie power parameters, respectively. The proposed model is equivalent to the exponential-Poisson-Tweedie models arising from geometric sums of Poisson-Tweedie random variables. In this respect, the PET models encompass the geometric versions of Hermite, Neyman Type A, P\'{o}lya-Aeppli, negative binomial and Poisson inverse Gaussian models. The algorithms we shall propose allow us to estimate the real power parameter, which works as an automatic distribution selection. Instead of the classical Poisson, zero-shifted geometric is presented as the reference count distribution. Practical properties are incorporated into the PET of new relative indexes of dispersion and zero-inflation phenomena. Simulation studies demonstrate that the proposed model highlights unbiased and consistent estimators for large samples. Illustrative practical applications are analysed on count datasets; in particular, PET models for data without covariates and PET regression models. The PET models are compared to Poisson-Tweedie models showing that parameters of both models are adopted to data.
\end{abstract}

\end{frontmatter}
\section{Introduction}\label{Int}
In the analysis of the count data, the presence of at least overdispersion or excess of zeros deserves special attention for the choice of the count model (e.g., Dem\'etrio and Hinde 1998; Akantziliotou et al. 2008; Sellers and Raim 2016; del Castillo and P\'{e}rez-Casany 2005). The overdispersion phenomenon which implies the fact that the variance (observed) is greater than the mean (expected variance) can also be induced by the zero-inflation one (or excess of zeros) in the sample or the heavy tail. Both measures are commonly performed with respect to the Poisson distribution and, frequently, the negative binomial and its zero-inflated version are used for modelling these kinds of count datasets. The phenomenon of overdispersion, often results from unobserved heterogeneity, i.e., the sample of responses is drawn from a population consisting of several sub-populations. Many count models have been built through compounding and mixing Poisson or negative binomial distributions; see, e.g., Hougaard et al. (1997), Kokonendji et al. (2004; 2007), Bonat et al. (2018), Wang (2011) and Gupta et al. (2014). The basic objective of this paper is to provide answers to the following three questions. What to do when the degree or level of overdispersion is very high? Should we relativize its measure with respect to another reference count distribution apart from Poisson? How to built a new family of ultra-overdispersed count models?


Consider for instance the number of occurrences of repairs for 2 549 buildings (e.g., plumbing, roof, heating/cooling system, etc.) over two years 1982 and 1983. These buildings were erected between 1945 and 1955 and were maintained by the Department of Engineering and Housing. The age of the structure influences the number of occurrences of repairs. Thus, higher maintenance actions may occur on a building than any other building in the entire database. The source of data is Yeoeman (1987). Being ultra-overdispersed data with sample variance equal to $20350.350$ and sample mean  amounting to $61.913$ stands for a crucial motive to explore a new class of models which can be called geometric-Poisson-Tweedie.

Analogously to the exponential dispersion models and their variance functions (J{\o}rgensen 1997), geometric and discrete dispersion models have been introduced later by J{\o}rgensen and Kokonendji (2011; 2016) as dispersion models for geometric sums and for count variables, respectively. It is attractive to write the random variable $Y$ as the geometric sum of discrete variables
\begin{equation}\label{geom}
Y=\sum_{\ell=1}^{G}\mathrm{PT}_{\ell}
\end{equation}
where $\mathrm{PT}_{1}, \mathrm{PT}_{2},\dots$ are independent and identically distributed (i.i.d.) such as $\mathrm{PT}$, a given Poisson-Tweedie random variable (e.g., Kokonendji et al. 2004; Bonat et al. 2018), and $G$ is a geometric random variable,  independent of $\mathrm{PT}$, with probability mass function $\mathbb{P}(G=g)=q(1-q)^{g-1}$ for $q\in(0,1]$ and $g=1,2,\dots$; sse, e.g., Kalashnikov (1997). The geometric sums of random independent count variables appear in various fields of probability, especially in risk, queueing and reliability theory. Some distributional properties of discrete geometric sums have been reported in literature. For example, Shanthikumar (1988) proved that the failure rate of a discrete geometric sum is decreases if $\mathrm{PT}_{1}$ has a decreasing failure rate. Abid et al. (2019b) characterized geometric sums of Tweedie models for analyzing both continuous and semicontinuous data. The representation (\ref{geom}), also viewed as an exponential mixture (Abid et al. 2019a) of ordinary PT models, leads to call such models exponential-Poisson-Tweedie or Poisson-exponential-Tweedie (PET) pointing out that this kind of models deals with count data. Based on both geometric and discrete dispersion models, the class of PET has variance
\begin{equation}\label{Varm}
\mathrm{Var}Y =m+m^2+\phi m^p,
\end{equation}
where $m=\mathbb{E}Y$, $\phi$ and $p$ are the dispersion and Tweedie power parameters, respectively.

The Fisher dispersion index (Fisher 1934) or P-dispersion index, defined with respect to the Poisson distribution as the ratio of variance to mean, makes possible to classify distributions and make inferences. Being ultra-overdispersed with unlimited range $\mathbb{N}:=\{0,1,2,\dots\}$, the Fisher index leads to an extra dispersion and such an index may be uninterpretable. Since we are concerned with a different type of random phenomenon, an alternative reference model may be appropriate. From (\ref{Varm}), the dispersion phenomenon with respect to the negative binomial distribution with unit dispersion parameter that is the zero-shifted geometric distribution ($\hbox{G}_{0}$) is a natural choice to reduce the extra overdispersion. Note that the binomial index for distributions with finite range $\{0, \dots,N\}$ with fixed $N\in\mathbb{N}$ was already discussed by Engel and te Brake (1993); see also Weiss (2018, p. 15). More generally, we can refer to Kokonendji and Puig (2018) for multivariate relative indexes. In this context, we will introduce the $\hbox{G}_{0}$-dispersion index and the $\hbox{G}_{0}$-zero inflation index similar to the P-dispersion index and the P-zero inflation index (e.g., Puig and Valero 2006), respectively, to the evaluation the departure from $\hbox{G}_{0}$.

The three parameter Poisson-Tweedie models are Poisson Tweedie mixtures. Indeed, let $\mathrm{PT}\sim \mathrm{PTw}_{p}(\widetilde{m},\widetilde{\phi})$ with $\mathrm{PTw}_{p}(\widetilde{m},\widetilde{\phi})$ denotes the Poisson-Tweedie distribution with mean $\widetilde{m}$ and mean-variance relationship $\mathrm{Var PT}=\widetilde{m}+\widetilde{\phi}\widetilde{m}^{p}$, such that $\widetilde{\phi}>0$ and $p \in \{0\} \cup [1,\infty)$ (see, e.g. Hougaard et al. 1997; J{\o}rgensen 1997, p. 165-170;
Kokonendji et al. 2004; Zhu and Joe 2009). The index $p$ uniquely determines a model in the PT family. However, its probability mass function has no closed-form expression apart from the special case corresponding to the negative binomial distribution, obtained when $p = 2$, i.e. a Poisson gamma mixture. Further special cases include the Hermite ($p = 0$), Poisson positive stable ($p > 2$) and Poisson-inverse Gaussian ($p = 3$) distributions. Therefore, the model can easily handle count data in a unified way (see, e.g., Bonat et al. 2018). Although they are flexible models, PT are not suitable to treat data having both properties (\ref{geom}) and (\ref{Varm}). If one intends to use PT models in the context of ultra-overdispersion, PT model compensates the ultra-overdispersion through selecting a high dispersion parameter $\widetilde{\phi}$. In contrast of Tweedie models which are non suitable to treat data having properties of geometric Tweedie models araising form geometric sums of Tweedie models (see, e.g., Abid et al. 2019b), the choice of power and dispersion parameters of the flexible PET and PT models are adopted directly to data.

Introducing in Abid et al. (2019a) and following Abid et al. (2019b) for (semi-)continuous regression models, this paper deals with the count regression models in presence of ultra-overdispersion phenomenon. This class provides a unified framework to deal essentially with ultra-overdispersed count data in the sense that, through the estimation of $p$, the obtained model automatically adapts both $\hbox{G}_{0}$-dispersion and $\hbox{G}_{0}$-zero-inflation indexes. As for the regression modeling, the proposed class is specified using only second-moments assumptions, that is variance and expectation for estimation and inference and may be fitted using the estimating function approach. In spite of the $\hbox{G}_{0}$-overdispersion of the PET model, the estimating function approach allows us to extend PETs to deal with $\hbox{G}_{0}$-underdispersed count data by allowing negative values of $\phi$.

The rest of the paper is organized as follows. In Section \ref{Sec2}, we firstly introduce and describe the PET family of distributions for $p\geq1$. Then, we investigate properties of both $\hbox{G}_{0}$-dispersion and $\hbox{G}_{0}$-zero-inflation indexes comparing them to the classical indexes.
In Section \ref{Sec3}, we identify the PET regression models, set forward the estimating function approach for parameters estimation and present some simulation studies. In Section \ref{Sec4}, we analyse interesting applications of the considered model in the insurance and reliability fields. In Section \ref{Sec5}, we conclude with some remarks and future directions of works.

\section{Poisson-exponential-Tweedie (PET) models}\label{Sec2}

\subsection{Definitions and interpretations}

We have now investigated the main properties of Poisson-exponential-Tweedie (PET) models, taking into consideration its mixture equivalent model. We are, as a matter of fact, imbedding the several distributions of interest.

The PET family is expressed by the following hierarchical formulation
\begin{equation}\label{PGT}
Y|Z \sim \mathrm{Poisson}(Z),~~
Z  \sim \mathrm{Tw}_{p}(Xm,X^{1-p}\phi)~~\mathrm{and}~~
X  \sim \mathrm{Exp}(1).
\end{equation}
where $\mathrm{Exp}(\lambda)$, $\mathrm{Poisson}(\lambda)$ and $\mathrm{Tw}_{p}(\lambda,\psi)$ denote the exponential distribution with parameter $\lambda$, the Poisson distribution with parameter $\lambda$ and the Tweedie distribution with mean $\lambda$, dispersion parameter $\phi$ and Tweedie power parameter $p$, respectively. In this case the PET is a factorial dispersion model. The geometric versions (\ref{geom}) of PT class collapses to an exponential mixture representation (Abid et al. 2018a, Proposition 2.4) that is expressed as
\begin{equation}\label{GPT}
X \sim \mathrm{Exp}(1),~~
[Y|X]|Z  \sim \mathrm{Poisson}(Z)~~\mathrm{and}~~
Z \sim \mathrm{Tw}_{p}(Xm,X^{1-p}\phi).
\end{equation}
For both models (\ref{GPT}) and (\ref{PGT}), we require $p\geq1$ to ensure $Z$ is non negative. The following proposition demonstrates the fact that two different random processes generate the same probability model and we need to explore it in the remaining of the paper so as to compute approximating likelihood.

\begin{prop}\label{dist}
Let $Y_{1}$ and $Y_{2}$ be two random variables defined by (\ref{PGT}) and (\ref{GPT}), respectively. Then:
\begin{enumerate}
[(i)] \item $Y_{1}$ and $Y_{2}$ have the same distributions given by
\begin{equation}\label{densite}
\!\mathbb{P}(Y_{1}=y)=\int_{0}^{\infty}\int_{0}^{\infty}\frac{\exp\{-z-x\}z^{y}}{y!}\mathrm{Tw}_{p}(mx,\phi x^{1-p})(z)dzdx=\mathbb{P}(Y_{2}=y).\end{equation}
 \item We have (\ref{Varm}) for $Y=Y_{1}=Y_{2}$.
\end{enumerate}
\end{prop}
\textbf{Proof:} (i) The probability mass function of $Y_{1}$ and $Y_{2}$ drawn from (\ref{PGT}) and (\ref{GPT}), respectively can be expressed as (\ref{densite}).\\
(ii) Decomposing $Y_{1}$ with respect to the Tweedie random variable $Z$, one has
$\mathrm{Var}Y_{1}=\mathbb{E}\{\mathrm{Var}(Y_{1}|Z)\}+\mathrm{Var}\{\mathbb{E}(Y_{1}|Z)\}
=\mathbb{E}Z+\mathrm{Var}Z$ and then the result is deduced. \hfill \qed

Based on this result, the PET distribution may be interpreted as an exponential mixture of PT distribution. This class is denoted $\mathrm{PETw}_{p}(m,\phi)$ for $p\in\{0\}\cup(1,\infty)$ and $m\in(0,\infty)$. A special case includes the negative binomial distribution obtained when $p=2$ with variance-mean relationship $m+(1+\phi)m^2$. Furthermore, a common member of PT, Hinde-Dem\'{e}trio and PET classes stands for the negative binomial distribution when $p=2$.

\subsection{Properties}

Computing index for P-dispersion, with respect to the Poisson distribution, are generally a first step in real count analysis, to explore the flexibility of the model. Another characteristic index is P-zero-inflation which is defined as the proportion of observing zeros compared to the Poisson distribution.
\begin{equation}\label{Poisson-ind}
\hbox{P-DI}=\frac{\hbox{Var}Y}{\mathbb{E}Y}~~~ \hbox{and}~~~\hbox{P-ZI}=\mathbb{E}Y + \log \mathbb{P}(Y=0).
\end{equation}
Values for $\hbox{P-DI}$ deviating from $1$ express a violation of the Poisson model. The P-DI indicates P-overdispersion for P-DI $>1$, P-underdispersion for P-DI $<1$ and P-equidispersion for P-DI $=1$. The corresponding centered dispersion index can be defined as $(\hbox{Var}Y-\mathbb{E}Y)/\mathbb{E}Y$. The definition of $\hbox{P-ZI}$ considered in this paper is slightly different from the familiar one reported in literature which is expressed by $\hbox{P-ZI}=1+\log \mathbb{P}(Y=0)/\mathbb{E}Y$. However, the one we considered is useful for defining relative zero-inflation index later. The P-zero-inflation indicates P-zero-inflation for P-ZI $>0$, P-zero-deflation for P-ZI $<0$ and P-no excess of zeros for P-ZI $=0$. Yet, since we are concerned with an ultra-overdispersion phenomenon (Part (ii) of Proposition \ref{dist}), it is so natural to consider the unit negative binomial that is the zero-shifted geometric ($\hbox{G}_{0}$) as an alternative reference model. Indeed, the $\hbox{G}_{0}$-dispersion ($\hbox{G}_{0}$-DI) and $\hbox{G}_{0}$-zero-inflation ($\hbox{G}_{0}$-ZI) indexes are defined as measures of departure from the $\hbox{G}_{0}$ model by
\begin{equation}\label{nb-ind}
\hbox{$\hbox{G}_{0}$-DI}=\frac{\mathrm{Var}Y}{\mathbb{E}Y+(\mathbb{E}Y)^2}~~~ \hbox{and}~~~\hbox{$\hbox{G}_{0}$-ZI}=\log(1+\mathbb{E}Y)+\log \mathbb{P}(Y=0).
\end{equation}
In view of these indexes, the $\hbox{G}_{0}$ distribution always satisfies $\hbox{$\hbox{G}_{0}$-DI}=1$, while a distribution
with $\hbox{$\hbox{G}_{0}$-DI}>1$ is said to exhibit $\hbox{G}_{0}$-overdispersion and $0<\hbox{$\hbox{G}_{0}$-DI}<1$ indicates $\hbox{G}_{0}$-underdispersion. Note that $\hbox{$\hbox{G}_{0}$-ZI}$ takes values in $(0,\infty)$. Furthermore, $\hbox{$\hbox{G}_{0}$-ZI}$ takes the value $0$ for the $\hbox{G}_{0}$ distribution, but may differ otherwise. Values $\hbox{$\hbox{G}_{0}$-ZI}>0$ express $\hbox{G}_{0}$-zero-inflation, while $\hbox{$\hbox{G}_{0}$-ZI}<0$ refers to $\hbox{G}_{0}$-zero-deflation. If we characterize the $\hbox{G}_{0}$ distribution behavior in terms of the Poisson indexes, it is P-overdispersed and P-zero inflated, while the Poisson distribution is underdispersed and zero deflated with respect to the $\hbox{G}_{0}$ distribution.
\begin{prop}
PET is overdispersed with relation to P and $\hbox{G}_{0}$, respectively.
\end{prop}
\textbf{Proof:} From Proposition \ref{dist}, we deduce $\mathrm{Var}Y>\mathbb{E}Y+\mathbb{E}Y^2$ and therefore the overdispersion with respect to the Poisson and $\hbox{G}_{0}$ distributions. \hfill \qed

The calculation of $\mathbb{P}(Y=0)$ requires the implementation of the probability mass function of the PET distribution (\ref{densite}) which is analytically tractable. The approximation of the double integral is based on the Monte Carlo integration and independent Tweedie simulations through the package \texttt{tweedie} (Dunn 2013) for the statistical software \texttt{R} (R Core Team, 2018). The number of random simulations affect the numerical stability in such a complex model. A random sample of at least size $10^{6}$ is required to have a reasonable approximation. Instead, Gauss-Laguerre method is an alternative integral approximation.

\begin{figure}[tph]
\centering
\setlength\fboxsep{6pt}
\setlength\fboxrule{0pt}
    ~~\includegraphics[width=0.86\textwidth]{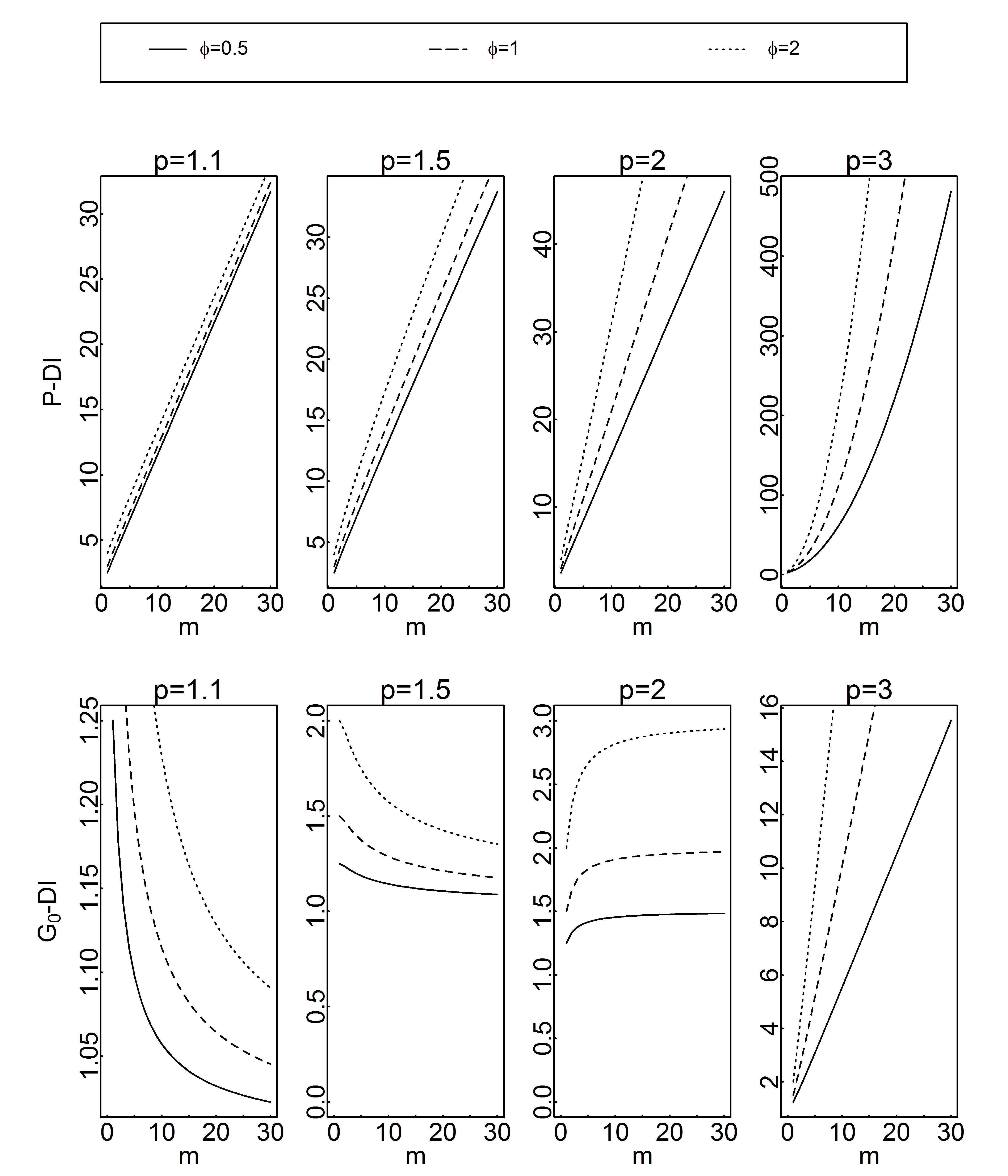}
   \caption{Dispersion indexes for PET distribution as a function of $m$ by dispersion and Tweedie power parameter values.}
\label{disp}
\end{figure}

\begin{figure}[tph]
\centering
\setlength\fboxsep{6pt}
\setlength\fboxrule{0pt}
    ~~\includegraphics[width=0.86\textwidth]{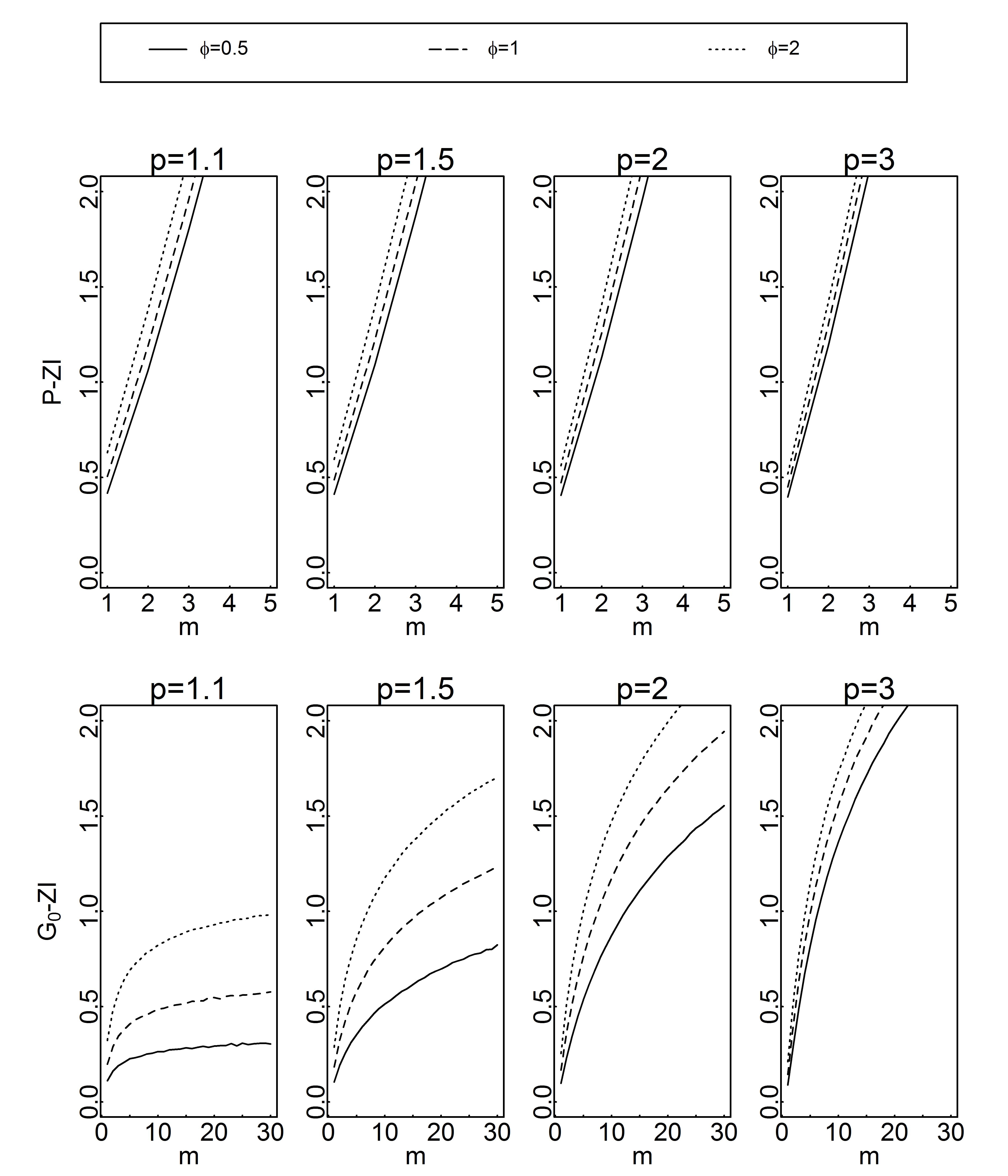}
   \caption{zero-inflation indexes for PET distribution as a function of $m$ by dispersion and Tweedie power parameter values.}
\label{zero}
\end{figure}

Figure \ref{disp} presents the P-dispersion and $\hbox{G}_{0}$-dispersion indexes as a function of the mean for different values of the power and dispersion parameters while Figure \ref{zero} illustrates P-zero-inflation and $\hbox{G}_{0}$-zero-inflation indexes for the same Tweedie and dispersion parameters values. The indexes relative to the Poisson distribution P-DI and P-ZI are slightly independent of the dispersion and power parameters and the curves are quite similar and indistinguishable for all scenarios. In general, the P-DI and P-ZI increase quickly as the mean increases giving an extremely overdispersed and zero inflated model for large values of the mean. It can be inferred from Figure \ref{disp} that the largest P-DI is $500$. Figure \ref{zero} depicts that like with small values of the mean, the P-ZI is equally important.
As can be noticed, the indexes related to the Poisson model are insignificant and can not be explained. In contrast, the difference of $\hbox{G}_{0}$ indexes for different scenarios is spectacular. The $\hbox{G}_{0}$-DI is dependent only on the values of the Tweedie power parameter distinguishing each distribution of PET models. The distribution is, of course, suitable to deal with ultra-overdispersed counts. For small values of the power parameter, the index $\hbox{G}_{0}$-DI tends to stabilize for large values of $m$. Meanwhile, for large values of the power parameter, the indexes increase rapidly with increasing mean, showing an extremely $\hbox{G}_{0}$-overdispersed model.

\begin{figure}[tph]
\centering
\setlength\fboxsep{6pt}
\setlength\fboxrule{0pt}
    ~~\includegraphics[width=0.86\textwidth]{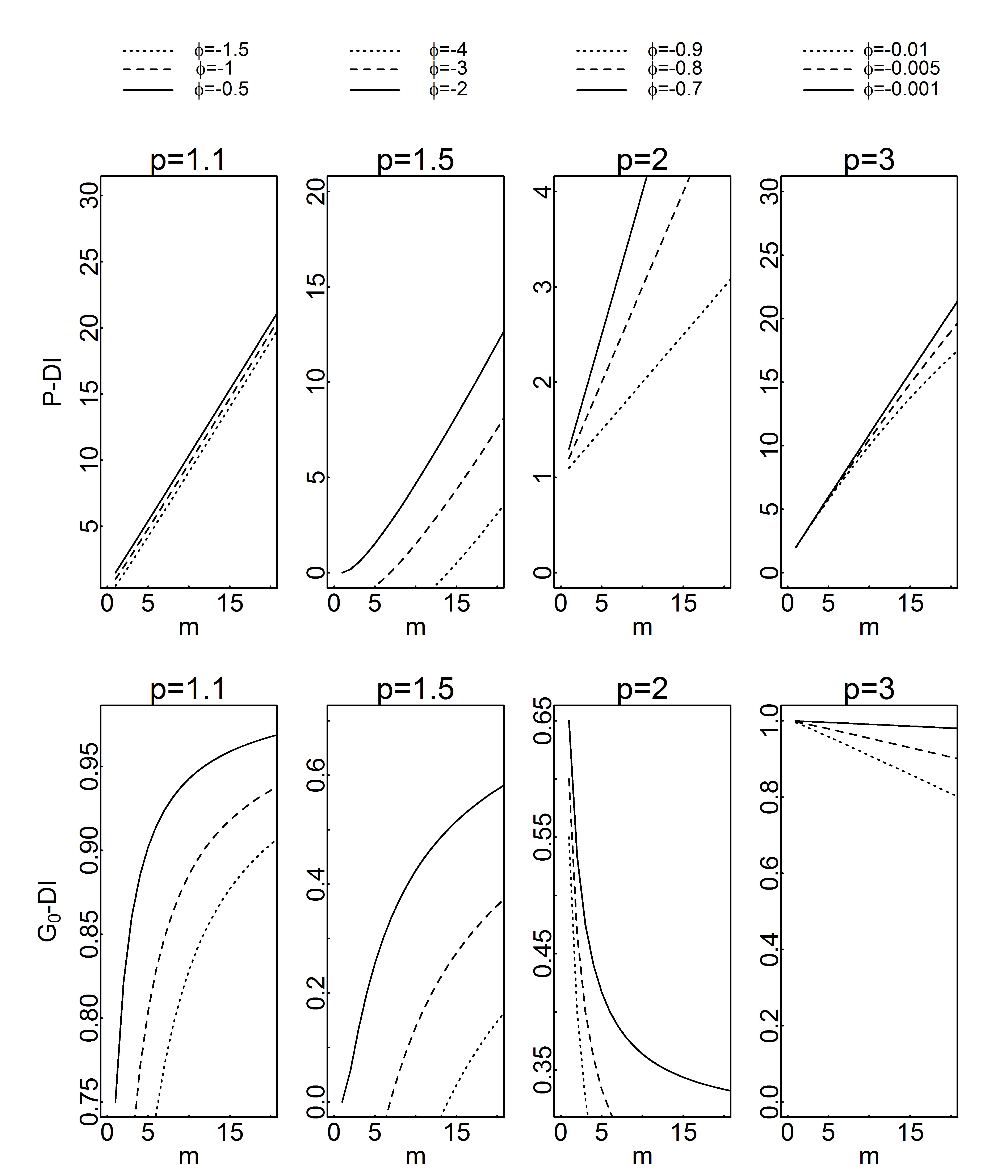}
   \caption{Dispersion indexes for PET distribution as a function of $m$ by different values of the
Tweedie power parameter and negative values of the dispersion parameter.}
\label{PDI_n}
\end{figure}

Figure \ref{PDI_n} exhibits the dispersion index as a function of the mean $m$ relative to the Poisson and $\hbox{G}_{0}$ distributions, respectively. We consider different values of the Tweedie power parameter and negative values for the dispersion parameter $\phi$. As it needs to be, for negative values of the dispersion parameter, the $\hbox{G}_{0}$-DI provides values smaller than $1$, indicating $\hbox{G}_{0}$-underdispersion while the P-DI provides values bigger than $1$, indicating P-overdispersion. Moreover, we also observe that, as the mean increases, the $\hbox{G}_{0}$-DI increases
normally for the Tweedie power parameter between 1 and 2. In addition, as the mean increases, the $\hbox{G}_{0}$-DI decreases rapidly for $p=2$. However, the $\hbox{G}_{0}$-DI decreases very slowly in such a way that the $\hbox{G}_{0}$-underdispersion becomes so negligible and insignificant for larger values of
the Tweedie power parameter. This demonstrates that the range of negative values allowed for the dispersion parameter decreases rapidly as the value of the Tweedie power parameter increases. Thus, for $\hbox{G}_{0}$-underdispersion data,
small and negative values are expected for the Tweedie power parameter.
Moreover, the heavy tail (HT) index is independent of the reference model and is given by
\begin{equation*}
\hbox{HT}=\frac{\mathbb{P}(Y=y+1)}{\mathbb{P}(Y=y)} ~~\hbox{for}~~y\rightarrow\infty.
\end{equation*}
Bonat et al. (2018) explored the properties of heavy tail index and asserted that for $p>2$, the PT model can deal with heavy tailed data. The PET models present the same behavior as PT models. Table \ref{Table_pGDMs} presents the main special cases and the dominant features of the PET model according to the values of the dispersion and Tweedie power parameters in relation to the Poisson and the $\hbox{G}_{0}$ distributions.
\begin{table}[tbh]
\begin{center}
\caption{Reference models and dominant features by dispersion and power parameter values in respect to the Poisson and $\hbox{G}_{0}$ models. ZI and HT stand for zero-inflation and heavy tail, respectively.}
\begin{tabular}{llll}
  \hline
   Reference & Power & Dominant features & Dispersion   \\
   \hline
   Poisson/$\hbox{G}_{0}$& - &Equi/Equi &$-$\\
   Geometric Hermite& $p=0$ &  Over, under &$\phi\lessgtr$0\\
   $[$Do not exist$]$&$0<p<1$&$-$& $-$ \\
   Geometric Neyman Type A&  $p=1$&Over, under, ZI &$\phi\lessgtr0$\\
   Geometric Poisson compound Poisson& $1<p<2$&Over, under, ZI&$\phi\lessgtr0$\\
   \emph{Geometric P\'{o}lya-Aeppli}&$p=1.5$&Over, under, ZI&$\phi\lessgtr0$\\
   Negative binomial&$p=2$&Over, under&$\phi\lessgtr0$\\
   Geometric Poisson positive stable &$p>2$&Over, HT&$\phi>0$\\
   \emph{Geometric Poisson-inverse Gaussian}&$p=3$&Over, HT&$\phi>0$\\
 \hline
\end{tabular}\label{Table_pGDMs}
    \end{center}
\end{table}
The hypothesis of the unilateral statistical test, in relation to the $\hbox{G}_{0}$ dispersion index will be expressed as
\begin{equation*}
\mathrm{H}_{0}:\hbox{G}_{0}\hbox{-DI}\leq1 \;\;\;\;\mathrm{versus}\;\;\;\; \mathrm{H}_{1}:\hbox{G}_{0}\hbox{-DI}>1.
\end{equation*}
When we reject the null hypothesis, the considered data are ultra-overdispersed with respect to the Poisson distribution (or more precisely, overdispersed with respect to the $\mathrm{G}_{0}$ distribution). Both count distributions PET and Poisson-Tweedie with $p>2$ are appropriate in such a situation. See, e.g., Miz\`{e}re et al. (2006) for testing the adequacy of the Poisson with counts data against alternatives of overdispersion and underdispersion.
\section{PET regression models}\label{Sec3}

Consider a cross-sectional dataset, $(y_{i},\textbf{x}_{i}),
$ $i=1,\dots,n,$ where $y_{i}$s are i.i.d realizations of $Y_{i}$ according to unspecified distribution $Y_{i}\sim\mathrm{PETw}_{p}(m_{i},\phi)$ and $\mathbf{x}_{i}$ is a $(q\times1)$ vector of known covariates.

Thus, because $m_{i}>0$, it is natural to model
\begin{equation}\label{esperance}
\mathbb{E}Y_{i}=m_{i}=\exp(\mathbf{x}_{i}^\top \boldsymbol{\beta})\\
\end{equation}
\begin{equation}\label{variance}
\mathrm{Var}Y_{i}=m_{i}+m_{i}^2+\phi m_{i}^{p},
\end{equation}
where $\boldsymbol{\beta}$ is a vector of unknown regression coefficients. In the meantime, the PET regression model is parametrized by $\boldsymbol{\theta}=(\boldsymbol{\beta}^\top,\boldsymbol{\gamma}^\top)^\top$, with $\boldsymbol{\gamma}=(p,\phi)$.

All PET regression models are designed, of course, to process $\hbox{G}_{0}$-overdispersed data for $\phi>0$ in (\ref{variance}). The only restriction for having a significant model is that $V_{i}>0$, that is,
\begin{equation*}\label{phi}
\phi>-m^{2-p}-m^{1-p}.
\end{equation*}
From this perspective, negative values for the dispersion parameter can also be required. As a consequence, the PET models may be extended to deal with $\hbox{G}_{0}$-underdispersed data. However, the associated density functions do not exist for $\phi <0$. Nevertheless, in a regression modelling framework, we are basically interested in the regression coefficient effects. As a matter of fact, such an issue does not imply any loss of interpretation and applicability.

\subsection{Estimation and inference}\label{S6ex}

We shall now present the estimating function approach using terminology and results from J{\o}rgensen and Knudsen (2004) and Bonat and J{\o}rgensen (2018). The quasi-score and Pearson estimating functions are adopted for estimation of regression and dispersion parameters, respectively. The quasi-score function for $\boldsymbol{\beta}$ has the following form:
\begin{equation}\label{quasi}
\psi_{\boldsymbol{\beta}}(\boldsymbol{\beta},\boldsymbol{\gamma})=\left(\sum_{i=1}^{n} \frac{\partial{m_{i}}}{\partial{\beta_{1}}}V_{i}^{-1}(y_{i}-m_{i}),\dots,\sum_{i=1}^{n} \frac{\partial{m_{i}}}{\partial{\beta_{q}}} V_{i}^{-1}(y_{i}-m_{i}) \right)^\top,
\end{equation}
Straightforward calculations demonstrate that $\partial m_{i}/\partial \beta_{j}= m_{i}$ for $j=1,\dots,q$.
The entry $(j, k)$ of the $q \times q$ sensitivity matrix for $\psi_{\boldsymbol{\beta}}$ is provided by
$$S_{\beta_{jk}}=\mathbb{E}\frac{\partial}{\partial \beta_{k}}\psi_{\beta_{j}}(\boldsymbol{\beta},\boldsymbol{\gamma})=-\sum_{i=1}^{n} m_{i} x_{ij} V_{i}^{-1} x_{ik} m_{i}.$$
Similarly, the entry $(j, k)$ of the $q \times q$ variability matrix for $\psi_{\boldsymbol{\beta}}$ is indicated by
$$V_{\beta_{jk}}= \hbox{Cov}( \psi_{\beta_{j}}(\boldsymbol{\beta},\boldsymbol{\gamma}),\psi_{\beta_{k}}(\boldsymbol{\beta},\boldsymbol{\gamma}))=\sum_{i=1}^{n} m_{i} x_{ij} V_{i}^{-1} x_{ik} m_{i},$$
whose inverse $V_{i}^{-1}$ yields the asymptotic variance of the quasi-likelihood estimator for $\beta$. The Pearson estimating function for the component $\boldsymbol{\gamma}$ has the following general form
\begin{equation}\label{PEF}
\psi_{\boldsymbol{\gamma}}(\boldsymbol{\beta},\boldsymbol{\gamma})=\left[-\sum_{i=1}^{n} \frac{\partial V_{i}^{-1}}{\partial \phi}\{(y_{i}-m_{i})^{2}-V_{i}\},-\sum_{i=1}^{n} \frac{\partial V_{i}^{-1}}{\partial p}\{(y_{i}-m_{i})^{2}-V_{i}\}\right],
\end{equation}
which is an unbiased estimating function for $\boldsymbol{\gamma}$ based on the squared residuals $(Y_{i} - m_{i})^2$ with mean $V_{i}$. The entry $(j, k)$ of the $2 \times 2$ sensitivity matrix for the dispersion parameters is expressed by
$$S_{\gamma_{jk}}=\mathbb{E}\frac{\partial}{\partial \gamma_{k}}\psi_{\gamma_{j}}(\boldsymbol{\beta},\boldsymbol{\gamma}) =-\sum_{i=1}^{n} \frac{\partial V_{i}^{-1}}{\partial \gamma_{j}}V_{i}\frac{\partial V_{i}^{-1}}{\partial \gamma_{k}}V_{i},$$
where $\gamma_{j}$ and $\gamma_{k}$ denote either $\phi$ or $p$, giving
\begin{equation}\label{Sensi}
S_{\boldsymbol{\gamma}}=\begin{pmatrix}
-\sum_{i=1}^{n}(m_{i}^{p}/V_{i})^{2} &  -\sum_{i=1}^{n} \{\phi m_{i}^{2p} \log(m_{i})\}/V_{i}^{2}\\
-\sum_{i=1}^{n} \{\phi m_{i}^{2p} \log(m_{i})\}/V_{i}^{2} & -\sum_{i=1}^{n} \{\phi m_{i}^{p} \log(m_{i})/V_{i}\}^{2}
\end{pmatrix}
.
\end{equation}

In a similar way, the cross entries of the sensitivity matrix are given by
$$S_{\beta_{j}\gamma_{k}}=\mathbb{E}\frac{\partial}{\partial \gamma_{k}}\psi_{\beta_{j}}(\boldsymbol{\beta},\boldsymbol{\gamma}) =0$$
and
$$S_{\gamma_{jk}}=\mathbb{E}\frac{\partial}{\partial \beta_{k}}\psi_{\gamma_{j}}(\boldsymbol{\beta},\boldsymbol{\gamma}) =-\sum_{i=1}^{n} \frac{\partial V_{i}^{-1}}{\partial \gamma_{j}}V_{i}\frac{\partial V_{i}^{-1}}{\partial \beta_{k}}V_{i}.$$
Denoting by $\boldsymbol{\theta}$ a vector of parameters with components $\boldsymbol{\beta}$ and $\boldsymbol{\gamma}$, the joint sensitivity matrix for $\boldsymbol{\theta}$ is given by
$$S_{\boldsymbol{\theta}}=\begin{pmatrix}
S_{\boldsymbol{\beta}} &  0\\
S_{\boldsymbol{\gamma\beta}} & S_{\boldsymbol{\gamma}}
\end{pmatrix}
.
$$
Likewise, the joint variability matrix of $\psi_{\boldsymbol{\beta}}$ and $\psi_{\boldsymbol{\gamma}}$ is given by
$$V_{\boldsymbol{\theta}}=\begin{pmatrix}
V_{\boldsymbol{\beta}} &  V_{\boldsymbol{\beta\gamma}}\\
V_{\boldsymbol{\gamma\beta}} & V_{\boldsymbol{\gamma}}
\end{pmatrix}
,
$$
where $V_{\boldsymbol{\beta\gamma}}=V_{\boldsymbol{\gamma\beta}}$ and $V_{\boldsymbol{\gamma}}$ depend on the third and fourth moments of $Y_{i}$, respectively. In order to avoid this dependence on high-order moments, we propose instead the
use of the empirical versions of $V_{\boldsymbol{\gamma}}$ and $V_{\boldsymbol{\gamma\beta}} $, which are given by
\begin{equation*}
\widetilde{V}_{\boldsymbol{\gamma}_{j,k}}(\boldsymbol{\beta},\boldsymbol{\gamma})=\sum_{i=1}^{n}\psi_{\boldsymbol{\gamma}_{j}}(\boldsymbol{\beta},\boldsymbol{\gamma})_{i}
\psi_{\boldsymbol{\gamma}_{k}}(\boldsymbol{\beta},\boldsymbol{\gamma})_{i}
~~~\hbox{and}~~~\widetilde{V}_{\boldsymbol{\gamma}_{j},\boldsymbol{\beta}_{k}}(\boldsymbol{\beta},\boldsymbol{\gamma})
=\sum_{i=1}^{n}\psi_{\boldsymbol{\gamma}_{j}}(\boldsymbol{\beta},\boldsymbol{\gamma})_{i}\psi_{\boldsymbol{\beta}_{k}}(\boldsymbol{\beta},\boldsymbol{\gamma})_{i}.
\end{equation*}
Let $\widehat{\boldsymbol{\theta}}$ denote the solution to the system of equations $\psi_{\boldsymbol{\gamma}}(\boldsymbol{\beta},\boldsymbol{\gamma})=0$ and $\psi_{\boldsymbol{\beta}}(\boldsymbol{\beta},\boldsymbol{\gamma})=0$. Then, the asymptotic distribution of $\boldsymbol{\theta}$ is $\widehat{\boldsymbol{\theta}}\sim N(\boldsymbol{\theta},J_{\boldsymbol{\theta}}^{-1})$ where $J_{\boldsymbol{\theta}}^{-1}$
is the inverse of Godambe information matrix,
$J_{\boldsymbol{\theta}}^{-1}=S_{\boldsymbol{\theta}}^{-1}V_{\boldsymbol{\theta}}S_{\boldsymbol{\theta}}^{-\top}$, with $S_{\boldsymbol{\theta}}^{-\top}=(S_{\boldsymbol{\theta}}^{-1})^{\top}$. J{\o}rgensen and Knudsen (2004) proposed the algorithm $$
\boldsymbol{\beta}^{(i+1)}=\boldsymbol{\beta}^{(i)}-S_{\boldsymbol{\beta}}^{-1}\psi_{\boldsymbol{\beta}}(\boldsymbol{\beta}^{(i)},\boldsymbol{\gamma}^{(i)})\;\;
\mathrm{and}\;\;
\boldsymbol{\gamma}^{(i+1)}=\boldsymbol{\gamma}^{(i)}-\alpha S_{\boldsymbol{\gamma}}^{-1}\psi_{\boldsymbol{\gamma}}(\boldsymbol{\beta}^{(i+1)},\boldsymbol{\gamma}^{(i)}),
$$
where $\alpha$ is a tuning constant used to control the step-length.

Note that the quasi-score function $\psi_{\boldsymbol{\beta}}(\boldsymbol{\beta},\boldsymbol{\gamma})$ is $\boldsymbol{\gamma}$-insensitive (J{\o}rgensen and Knudsen 2004) which allows us to use two separate equations
to update $\boldsymbol{\beta}$ and $\boldsymbol{\gamma}$. Furthermore, the modified chaser algorithm is guaranteed to converge under good initial values. Parameter estimates from the fit of a standard Poisson regression models are recommended to obtain values for the regression coefficients. The Pearson estimator is recommended to obtain the initial value of dispersion parameter.

\subsection{Simulation studies}

In this section, we shall perform a simulation study to assess the flexibility of the PET regression model to deal with ultra-overdispersed count data.

The expectation and the variance of the PET random variable are given by $$m_{i}= \exp(\beta_{0} +\beta_{1}x_{1i} +\beta_{2}x_{2i})~~~\hbox{and}~~~V_{i}=m_{i}+m_{i}^2+\phi m_{i}^{p},$$
where
$x_{1}$ and $x_{2}$ are sequences from $-1$ to $1$ with a length that is equal to the sample size. The regression coefficients were fixed at the values, $\beta_{0} = 1$, $\beta_{1} = -1$ and $\beta_{2}=-0.9$. Different sample sizes ($n=500$, $1000$ and $5000$) generating $1000$ data sets are used in each case, and two measures of estimator quality are compared (bias and coverage rate). In this manner, we have a quality measure based on point estimates and another one based on confidence intervals. Four values of the Tweedie power parameter $p=1.01,1.5,2,3$ are considered combined with three values of the dispersion parameter $\phi=0.5$, $1$, $1.5$. The limiting case $p = 0$ corresponds to the geometric Hermite distribution. The availability
of simulation procedures for Hermite distributions through the \texttt{hermite} package (Higueras et al. 2015) makes easy the simulation of the limiting case $p = 0$ corresponding to the geometric Hermite distribution. However, in such models we require $m_{i}>\phi$ and then only small values of the dispersion parameter $\phi$ may be allowed.
\begin{figure}[tph]
\centering
\setlength\fboxsep{6pt}
\setlength\fboxrule{0pt}
~~\includegraphics[width=1\textwidth]{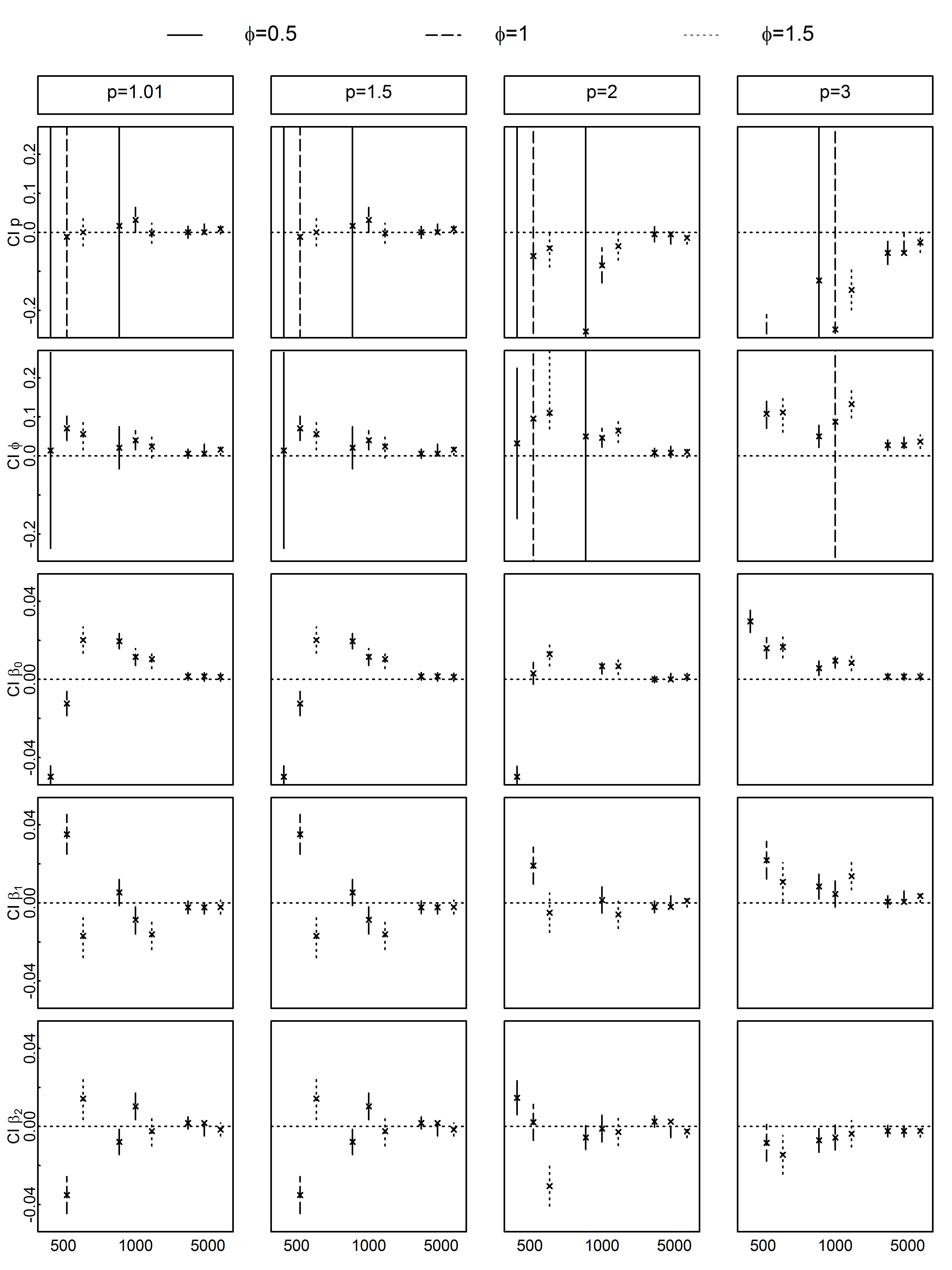}
   \caption{Average bias and confidence intervals for different sample sizes ($n=500$, $1000$, $5000$) and parameter combinations.}
\label{figicdisp}
\end{figure}
Figure \ref{figicdisp} presents the bias and confidence intervals for the regression and dispersion parameters under each scenario.
It can be inferred from Figure \ref{figicdisp} that for all scenarios, the bias tends to $0$ as the sample size increases.
These results illustrate the asymptotic consistency and unbiasedness of the estimating function estimators of the regression and dispersion coefficients. Note that, for large sample size, the largest bias for parameter estimates is equal to one percent. It is also noticeable that the bias of the estimators increases when the value of $p$ increases, which is expected because large values of $p$ entail large $\hbox{G}_{0}$-overdispersion. Thus, in these scenarios large samples are required to carry out a correct estimation. Furthermore, for sample size around $1000$, the bias of the quasi-score approach estimators is too small to be useful for practical situations. It is worthnoting that the confidence intervals based on the quasi-score approach are narrow, indicating that estimators are precise and the confidence intervals are well estimated. The worst results appear for the power parameter with large $\hbox{G}_{0}$-overdispersion.
Overall, the results demonstrate that the proposed algorithm leads to satisfactory estimates. However, this method requires large sample sizes, which is expected because the inferential methods are based on asymptotic results.

\section{Applications to real count data}\label{Sec4}

In this section, several instances of applications are displayed from real data. The first concerns an application of PET model without covariables while the two others illustrate the application of the new PET regression models.

\subsection{Accidents of private cars in Switzerland}

This example concerns the number of automobile liability policies in Switzerland for private cars taken from Klugman (2004). The estimated dispersion indexes $\widehat{\hbox{P-DI}}=1.154$ and $\widehat{\hbox{$\hbox{G}_{0}$-DI}}=0.999$ highlight the overdispersion with respect to the Poisson distribution and the quasi equidispersion with respect to the $\hbox{G}_{0}$ distribution. The values of the estimated zero-inflation indexes $\widehat{\hbox{P-ZI}}= 0.154$ and $\widehat{\hbox{$\hbox{G}_{0}$-ZI}}=-2.709$ show a P-zero inflation and $\hbox{G}_{0}$-zero deflation. This data set is analysed in Aryuyuen and Bodhisuwan (2013) using the negative binomial-generalized exponential (NB-GE) distribution with parameters $r$, $\alpha$, $\beta>0$ as an example of P-overdispersed data having excess of zeros. Following Aryuyuen and Bodhisuwan (2013), the mean of the $\hbox{G}_{0}$-GE is $r\delta_{1}-r$ and the variance is $-r^2\delta_{1}^2-r\delta_{1}+(r^2+r)\delta_{2}$, with $\delta_{1}=\Gamma(\alpha+1)\Gamma(1-\beta^{-1})/\Gamma(\alpha-\beta^{-1}+1)$ and $\delta_{2}=\Gamma(\alpha+1)\Gamma(1-2\beta^{-1})/\Gamma(\alpha-2\beta^{-1}+1)$. An experimental study was performed based on the approximated maximum likelihood approach to select parameters based on the PET and PT models with different values of $p\geq1$ and $\phi>0$. Table \ref{acci} presents comparison results among the Poisson, negative binomial, PT, negative binomial-generalized exponential and Poisson-exponential-Tweedie distributions.
\begin{table}[!]
\begin{center}
\caption{Parameter estimates and standard errors (SE) for PET and PT model (Aryuyuen and Bodhisuwan 2013); NB and NB-GE design negative-binomial and negative binomial-generalized exponential distributions, respectively; Chi-squares$^2$ denotes chi-square distribution with $2$ degrees of freedom.}
\begin{tabular}{crrrrrr}
\hline
\multicolumn{1}{c}{Number of} & \multicolumn{1}{c}{Observed}&\multicolumn{3}{c}{Fitting distributions}\\
accidents &frequencies&Poisson&$\mathrm{NB}\;\;\;\;$&NB-GE&$\mathrm{PT}\;\;\;\;\;$& $\mathrm{PET}\;\;\;\;$\\
 \hline
$0$ &$103704$&$102629.6$&$103723.6$&$103708.8$&$103708.1$&$103708.8$\\
$1$ &$14075$&15922.0&13989.9&14046.8&14041.3&14044.9\\
$2$ &$1766$&1235.1&1857.1&1797.8&1809.6&1797.3\\
$3$&$255$&$+\;66.3$&245.2&251.7&250.3&259.1\\
$4$ &$45$ &&$+\; 37.2$&36.0&36.9&36.6\\
$5$ &$6$&&&$+\; 12$&$+\; 6.8$&+$\;6.3$\\
$6$ &$2$&&&&&\\
$\;+7$&$0$&&&&&\\
\hline
Parameters&& $\widehat{\lambda}=0.155$&$\widehat{r}= 1.032$&$\widehat{r}= 2.431$&$\widehat{p}= 2.600$&$\widehat{p}= 1.950$\\
estimates &&&$\widehat{q}=0.150$&$\widehat{\alpha}= 3.289$&$\widehat{\phi}= 3.000$&$\widehat{\phi}= 0.050$\\
&&&&$\widehat{\beta}= 31.278$&$\widehat{m}= 0.155$&$\widehat{m}= 0.155$\\
\hline
Chi-squares$^2$&&$1332.300$&$12.120$&$4.260$&$3.209$& $2.932$\\
p-value&&$<0.0001$&$0.0023$&$0.1180$&$0.2010$&$0.2300$\\
\hline
\end{tabular}
\label{acci}
\end{center}
\end{table}

By comparing these fitting distributions in Table \ref{acci} and based on the p-value, the PET distribution provides a better fit than the PT and NB-GE distributions, even without additional covariates. A chi-square test also supports this statement. The estimated value of PT power parameter $\widehat{p}\geq2$ provides evidence of ultra-overdispersion that may be attributed to zero-inflation. In contrast, the PET detects the excess of zeros since the estimate power parameter is in $(1,2)$. Furthermore, it is interesting to note that the compensated PT model has a higher dispersion parameter $\widehat{\phi}= 3.000$ compared to that of PET model ($\widehat{\phi}= 0.050$).

\subsection{Accident occurrence in car insurance on Tunisian data}

This example is based on data that we received from an insurer who operates in the market for automobile insurance in Tunisia. The data consists of a cross section of 12 541 driving licence holders over the year 2015. It is noted that one registration corresponds to one covered risk/accident (if the policyholder has two cars, then he will do two registrations).
The main goal of the investigation was to assess the effect of the age of policyholder, the age and the power of the car which are factors of risk in car insurance. The estimated indexes are $\widehat{\hbox{P-DI}}=9.100$, $\widehat{\hbox{$\hbox{G}_{0}$-DI}}=6.204$,  $\widehat{\hbox{P-ZI}}=0.268$ and $\widehat{\hbox{$\hbox{G}_{0}$-ZI}}=0.185$. To compare the PET model to the PT model, Table \ref{tuni} exhibits the corresponding estimates and standard errors. A standard method of comparing count models is to use such information criteria, as the Akaike Information Criterion. As no analytic probability mass function exists, the pseudo version of this criterion is used.
\begin{table}[h]
\begin{center}
\caption{Parameter estimates and standard errors (SE) for PET and PT models; pAIC information criterion for models.}
\begin{tabular}{lrrrr}
\hline
Parameter & $\mathrm{PET}\;\;\;\;\;\;\;$ & $\mathrm{PT}\;\;\;\;\;\;\;$ \\
 \hline
Intercept & $0.267$ $(0.111)$ &$0.267$ $(0.111)$\\
Car age & $0.105$ $(0.003)$ & $0.105$ $(0.003)$\\
Car power& $0.112$ $(0.009)$ & $0.112$ $(0.009)$\\
Driver age & $-0.102$ $(0.002)$ & $-0.102$ $(0.002)$\\
$\phi$  & $0.529$ $(0.093)$ & $1.741$ $(0.096)$\\
$p$ & $2.840$ $(0.149)$ & $2.329$ $(0.075)$\\
\hline
pAIC &15287.230 & $15298.930$\\
\hline
\end{tabular}
\label{tuni}
\end{center}
\end{table}

The results presented in Table \ref{tuni} determin that the pAIC are quite similar for the PT and PET models. The small difference in terms of pAIC suggests the very competitive fit of the PET regression model. However, it is important to note that both models select power parameters estimate $\hat{p}>2$ since the degree of zero inflation is not high.
Furthermore, both models provide very similar regression estimates and standard errors, leading to
identical interpretations. The difference in the dispersion parameter estimate of both models PT and PET is spectacular since both models automatically adapt the dispersion in the data by the appropriate estimation of the dispersion parameter.

\subsection{Buildings maintenance data with double PETs}

The last dataset used to illustrate the methodology developed in this work is reported in Section \ref{Int}. Data of number of maintenance for all buildings are available during two years: $1982$, $1983$. It is assumed that buildings can be classified as repairable systems since they are repaired rather than thrown away each time a component breaks. Consider that, for a given year, the total number of buildings maintenance $Y_{i}$ in the $i$-th time frame follows the PET model $\mathrm{PETw}_p(m_{i},\phi)$, $i=1, \dots, 46$. First consider Table \ref{Indexes} presenting a summary of these four datasets corresponding to each year, along with the indexes P-DI, $\hbox{G}_{0}$-DI, P-ZI and $\hbox{G}_{0}$-ZI.
\begin{table}[tbh]
\begin{center}
\caption{Estimated dispersion and zero-inflation indexes of datasets.}
\begin{tabular}{lrrrr}
\hline
Dataset & $\widehat{\hbox{P-DI}}\;\;\:$ & $\widehat{\hbox{$\hbox{G}_{0}$-DI}}$ & $\widehat{\hbox{P-ZI}}\;\:$ & $\widehat{\hbox{$\hbox{G}_{0}$-ZI}}$\\
\hline
No $1$ (1982) & $328.692$&  $5.224$&$60.550$&$2.789$\\
No $2$ (1983) & $243.826$ & $7.393$&$30.539$&$2.232$\\
\hline
\end{tabular}
\label{Indexes}
\end{center}
\end{table}
To measure the departure from the Poisson and the $\hbox{G}_{0}$ distributions of the considered datasets, our estimated indexes provide a good summary of the dispersion and zero-inflation phenomenon. We detect a high degree of P-overdispersion and P-zero-inflation in all scenarios and consequently the Poisson and the PT are clearly unsuitable, being overly conservative. However, a weaker degree of both $\hbox{G}_{0}$-overdispersion and $\hbox{G}_{0}$-zero-inflation is detected pointing that $\hbox{G}_{0}$ indexes relativize very well the excess of overdispersion and zeroes. Indeed, $\widehat{\hbox{P-DI}}$ and $\widehat{\hbox{$\hbox{G}_{0}$-DI}}$ work in a different way in terms of the degree of overdispersion. This refers to the linear variance-mean relationship of P-DI and the quadratic variance-mean relationship of $\hbox{G}_{0}$-DI. The parameters estimates and goodness of fit measures for the PET and PT regression models are outlined in Table \ref{82-83} along with the pseudo Akaike information criteria.

\begin{table}[!]
\begin{center}
\caption{Parameter estimates and standard errors (SE) for PET and PT indicated by italic and roman symbols, respectively; pAIC for models.}
\begin{tabular}{lrr}
\hline
Parameter & No $1$ (1982) & No $2$ (1983)\\ 
\hline
\multirow{2}{*}{Intercept} & $\emph{3.494}$ $\emph{(0.440)}$&$\emph{2.984}$ $\emph{(0.342)}$\\
& 3.996 (0.299)&2.859 (0.477)\\
\multirow{2}{*}{Age}  & $\emph{0.078}$ $\emph{(0.010)}$&$\emph{0.015}$ $\emph{(0.019)}$\\
 & 0.080 (0.012) &0.023 (0.144)\\
\multirow{2}{*}{$\phi$}& $\emph{0.291}$ $\emph{(0.036)}$ &\emph{$2.749\;10^{-7}$} $\emph{(0.036)}$ \\
&0.691 (0.312) & $3.569\;10^{-4}$ (1.336)\\
\multirow{2}{*}{$p$} & $\emph{1.420}$ $\emph{(0.010)}$&$\emph{3.361}$ $\emph{(0.180)}$ \\
&2.010 (0.007) &4.656 (0.012)\\
\hline
\multirow{2}{*}{pAIC} &$\emph{3344.466}$&$\emph{542.176}$\\
 &3357.340&547.346\\
\hline
\end{tabular}
\label{82-83}
\end{center}
\end{table}

The results displayed in Table \ref{82-83} reveal small differences in terms of pAIC
values in favour of the PET regression model.
PT and PET models provide very similar regression estimates and standard errors but the dispersion and power estimates for both models are quite different which is analogous to the previous application. For the year 1982, the estimated power parameter $\widehat{p}>2$ of the PT model goes in good agreement with the ultra-overdispersion structure and does not highlight the presence of zeroes. However, the estimated power parameter $\widehat{p}\in(1,2)$ of the PET model with its corresponding standard error indicate that the PET can offer a very competitive fit to describe the excess of zeroes. The $\phi$ estimate of the PT model is $23 \%$ larger than that of PET model. The interpretations are similar to other cases studies. For the year 1983, this case is in particular challenging for both models.  In this case and for numerical reasons, the reparametrization $\phi=\exp\delta$ is a way to stabilize the fitting process (see, e.g., Abid et al. 2019b) for a similar approach. The estimated power parameter $\widehat{p}$ is high for both models to compensate small values of $\widehat{\phi}$.


\section{Concluding remarks}\label{Sec5}

In this paper, we have presented the novel PET distribution and its associated regression model. This model may be interpreted through both exponential-PT and Poisson-exponential-Tweedie mixtures. The main merit of the proposed model is that it offers a flexible statistical modelling framework to deal with ultra-overdispersed count data. The definition of the novel indexes $\hbox{G}_{0}$-DI and $\hbox{G}_{0}$-ZI set forward here, in which the $\hbox{G}_{0}$ appears to be the reference, proves to be easy and pertinent in terms of handling ultra-overdispersed data from a theoretical and practical point of view. Compared to the classical P-DI and P-ZI, the new indexes take into account the $\hbox{G}_{0}$-(over/under/equi)-dispersion and excess of zeros in relation to $\hbox{G}_{0}$. The interpretation and some properties of $\hbox{G}_{0}$-DI and $\hbox{G}_{0}$-ZI are provided.

We have afterwards introduced the PET regression models as regression modelling of PETs in the conventional framework of the generalized linear models. Furthermore, we adopted an estimating function approach for estimation and inference
based only on second-order moment assumptions. Such a specification allows us to extend the PT model to handle $\hbox{G}_{0}$-underdispersed count data by allowing negative values for the dispersion parameter. The estimation of the power parameter plays a significant role in the context of PET regression models, since it is an index that distinguishes important distributions and consequently can function as an automatic distribution selection.  Simulation studies showed that the proposed estimating function approach highlighted unbiased and consistent estimators for both regression and dispersion coefficients. In addition, we demonstrated through data
analysis that our model is quite flexible and can easily adapt to the suitable count distribution. Comparisons of PET with PT models show that PT model compensate the ultra-overdispersion by a higher estimate of the dispersion parameter.

At this stage of analysis, we would assert that our research is a step that may be taken further as
the basic model discussed in this paper is promising and may be extended in different ways. This involves the elaboration of multivariate ultra-overdispersed count data, with many potential applications for the analysis of longitudinal and spatial data. We may consider, for instance, the third data set as a prospective study, taking into account the correlation structure. Moreover, we also plan to extend the estimating function approach by regressing both of the mean and the dispersion parameter (e.g., Smyth 1989; Petterle et al. 2019). Finally, instead of the geometric compounding (\ref{geom}), the
negative binomial compounding can be considered (e.g., Iw\`{\i}nska and Szymkowiak 2017)
and a second dispersion parameter $\lambda > 0$ can be introduced and estimated using the moment method. These extensions will form the cornerstone of future works.

\section*{Acknowledgements}
The authors are particularly grateful to Wagner Bonat and Simplice Dossou-Gb\'{e}t\'{e} for numerous conversations during the preparation of this paper. Part of this work was supported by University of Sfax as it was performed while the second author was at the Laboratory of Probability and Statistics of Sfax.


\begin{thebibliography}{}
\bibitem{Ab2019a} Abid, R., Kokonendji, C.C., Masmoudi, A.: Geometric dispersion models with quadratic v-functions. Statist. Probab. Lett. 145, 197-204 (2019a)

\bibitem{Ab2019b} Abid, R., Kokonendji, C.C., Masmoudi, A.: Geometric Tweedie regression models for continuous and semicontinuous data with variation phenomenon. AStA Adv. in Statist. Anal. (2019b) DOI : 10.1007/s10182-019-00350-8

\bibitem{Ridd16} Akantziliotou, C., Rigby, R.A., Stasinopoulos, D.M.: A framework for
modelling overdispersed count data, including the Poisson-shifted generalized inverse Gaussian distribution. Comput. Statist. Data Anal. 53, 381-393 (2008)

\bibitem{AB13} Aryuyuen, S., Bodhisuwan, W.: The negative binomial-generalized exponential
(NB-GE) distribution. Appl. Math. Sci. 7, 1093-1105 (2013)


\bibitem{BWHJKHDMa2017} Bonat, W.H., J{\o}rgensen, B., Kokonendji, C.C., Hinde, J., Dem\'{e}trio, C.G.B.: Extended Poisson-Tweedie: properties and regression models for count data. Statist. Model. 18, 24-49 (2018)

\bibitem{delcastello205} del Castillo, J., P\'{e}rez-Casany, M.: Overdispersed and underdispersed Poisson generalizations. J. Statist. Plan. Infer. 134, 486-500 (2005)

\bibitem{Dunn2013} Dunn, P.K.: Tweedie exponential family models. version 2.1.7. R package URL \texttt{http://cran.r-project.org/web/packages/tweedie/tweedie} (2013)

\bibitem{BrEn93} Engel, B., te Brake, J.: Analysis of embryonic development with a model for under- or overdispersion relative to binomial variation. Biometrics 49, 269-279 (1993)

\bibitem{F34} Fisher, R.A.: The effects of methods of ascertainment upon the estimation of frequencies. Ann. Eug. 6, 13-25 (1934)
%
\bibitem{Gupt} Gupta, R.C., Sim, S.Z., Ong, S.H.: Analysis of discrete data by Conway–Maxwell Poisson distribution. AStA Adv. in Statist. Anal. 4, 327–343 (2014)

\bibitem{HMOP15} Higueras M., Moria D., Oliveira M., Puig P.: hermite: Generalized Hermite Distribution. R package URL \texttt{https://CRAN.R-project.org/package=hermite} (2015)
%
\bibitem{HD98} Hinde, J., Dem\'{e}trio, C.G.B.: Overdispersion: Models and Estimation. Associacao Brasileira de Estatistica, Sao Paulo (1998)

\bibitem{HLW15} Hougaard, P., Lee, M-L.T., Whitmore, G.A.: Analysis of overdispersed count data by mixtures of Poisson variables and Poisson processes. Biometrics 53, 1225-1238 (1997)
%
\bibitem{Iw17} Iw\`{\i}nska, M., Szymkowiak, M.: Characterizations of distributions through selected functions of reliability theory. Commun. Statist. Theo. Meth. 46, 69-74 (2017)

\bibitem{Jor97} J{\o}rgensen, B.: The Theory of Dispersion Models. Chapman and Hall, London (1997)

\bibitem{JorKn04} J{\o}rgensen, B., Knudsen, S.J.: Parameter orthogonality and bias adjustment for estimating functions. Scand. J. Statist. 31, 93-114 (2004)
%
\bibitem{JorKOk11} J{\o}rgensen, B., Kokonendji, C.C.: Dispersion models for geometric sums. Braz. J. Probab. Statist. 25, 263-293 (2011)
%
\bibitem{JorKok2016} J{\o}rgensen, B., Kokonendji, C.C.: Discrete dispersion models and their Tweedie asymptotics. AStA Adv. in Statist. Anal. 100, 43-78 (2016)

\bibitem{Kala97} Kalashnikov, V.: Geometric Sums: Bounds for Rare Events with Applications. Kluwer Academic, Dordrecht (1997)
%
\bibitem{Koko2014} Kokonendji, C.C.: Over- and underdispersion models. In The Wiley Encyclopedia of Clinical Trials - Methods and Applications of Statistics in Clinical Trials, pages 506-526 (Chapter. 30). Wiley, New York (2014)
%
\bibitem{KDZ07} Kokonendji, C.C., Dem\'{e}trio, C.G.B., Zocchi, S.S.: On Hinde-Dem\'{e}trio regression models for overdispersed count data. Statist. Meth. 4, 271-291 (2007)
%
\bibitem{KoDD04} Kokonendji, C.C., Dossou-Gbete, S., Dem\'{e}trio, C.G.B.: Some discrete exponential dispersion models: Poisson-Tweedie and Hinde-Demetrio classes. Statist. Oper. Res. Trans. 28, 201-214 (2004)

\bibitem{KoPU18} Kokonendji, C.C., Puig, P.: Fisher dispersion index for multivariate count distributions : A review and a new proposal. J. Mult. Anal. 165, 180-193 (2018)
%
\bibitem{Kl04} Klugman, S.A., Panger, H.H., Willmot, G.E.: Loss Models: From Data to Decisions. Wiley, Hoboken (2004)
%
\bibitem{MN89} McCullagh, P., Nelder, J.: Generalized linear models, 2nd edition. Chapman and Hall, London (1989)
%
\bibitem{Mkokd89} Miz\`{e}re, D., Kokonendji, C.C., Dossou-Gb\'{e}t\'{e}, S.: Quelques tests de la loi de Poisson contre des alternatives g\'{e}n\'{e}rales bas\'{e}s sur l'indice de dispersion de Fisher. Rev. Statist. Appliq. 54, 61-84 (2006)

\bibitem{P19} Petterle, R.R., Bonat, W.H., Kokonendji, C.C., Seganfredo, J.C., Moraes, A., Gomes da Silva, M.M.: Double Poisson-Tweedie regression models. Int. J. Biostatist. 15, DOI :10.1515/ijb-2018-0119 (2019)

\bibitem{P06} Puig, P., Valero, J.: Count data distributions: some characterizations with
applications. J. Am. Statist. Assoc. 101, 332-340 (2006)
%
\bibitem{R2016} R Core Team: R: A Language and Environment for
Statistical Computing. R Foundation for Statistical Computing,  Vienna (2018)
%
\bibitem{guuy1988} Sellers, K.F., Raim, A.: A flexible zero-inflated model to address data
dispersion. Comput. Statist. Data Anal. 99, 68-80 (2016)

%
\bibitem{Sh1988} Shanthikumar, J.G.: DFR property offirst passage times and its preservation under geometric compounding. Ann. Probab. 16, 397-406 (1988)
%
\bibitem{Tw1984} Tweedie, M.C.K.: An index which distinguishes between some important exponential families. In \textit{Statistics: Applications and New Directions. Proceedings of the Indian Statistical Institute Golden Jubilee International Conference} (J. K. Ghosh and J. Roy, eds.), pages 579-604. Indian Statistical Institute, Calcutta (1984)
%
\bibitem{W11} Wang, Z.: One mixed negative binomial distribution with application. J. Statist. Plan. Infer. 141, 1153-1160 (2011)

\bibitem{W18} Weiss, C.H.: An Introduction to Discrete-Valued Time Series. Wiley, Hoboken (2018)
%
\bibitem{Y18} Yeoeman, A.: Forecasting Building Maintenance Using The Weibull Process (M.S.Thesis). University of Missouri-Rolla, United States (1987)
%
\bibitem{z18} Zhu, R., Joe, H.: Modelling heavy-tailed count data using a generalized Poisson-inverse Gaussian family. Statist. Probab. Lett. 70, 1695-1703 (2009)
\end{thebibliography}
\end{document}